\documentclass[usenatbib,usegraphicx]{mn2e}

\usepackage{times} % times new roman font style
\usepackage{latexsym,amssymb} % packages for symbols
\usepackage{amsmath} % AMS-LaTeX package for e.g. subequations
\usepackage{journalnames} % translation of journal names in ADS bibitems (custom style file)
\usepackage{booktabs} % better table style

\newcommand{\mbh}{M_\bullet}
\newcommand{\Msun}{M_\odot}
\newcommand{\Lsun}{L_\odot}

\newcommand{\kpc}{\,\mathrm{kpc}}
\newcommand{\Mpc}{\,\mathrm{Mpc}}
\newcommand{\Gyr}{\,\mathrm{Gyr}}
\newcommand{\magarcsec}{\,\textrm{mag\,arcsec}^{-2}}
\newcommand{\Upsst}{\Upsilon_\star}
\newcommand{\Upssti}{{\Upsst}_{,i}}
\newcommand{\sigst}{\sigma_\star}
\newcommand{\kms}{\,\mathrm{km\,s^{-1}}}
\newcommand{\gfthree}{\textsc{Galfit3}}
\newcommand{\sersic}{S\'{e}rsic}
\newcommand{\vlos}{\mathrm{v_{los}}}
\newcommand{\Mst}{M_\star}
\newcommand{\Mdm}{M_\mathrm{DM}}
\newcommand{\miuscat}{\textsc{MIUSCAT}}

\title[Bottom-heavy initial mass function]{Bottom-heavy initial mass function in a nearby compact $L^\star$-galaxy}

\author[R. L\"{a}sker et al.]{Ronald L\"{a}sker$^1$, Remco C. E. van den Bosch$^1$, Glenn van de Ven$^1$, Ignacio Ferreras$^2$, \newauthor Francesco La Barbera$^3$, Alexandre Vazdekis$^4$ and Jes\'{u}s Falc\'{o}n-Barroso$^4$ \\
  ${}^1$Max Planck Institute for Astronomy, K\"onigstuhl 17, 69117 Heidelberg, Germany; eMail: laesker@mpia-hd.mpg.de \\
  ${}^2$Department of Space and Climate Physics, Mullard Space Science Laboratory, University College London, \\  Holmbury St Mary, Dorking, Surrey RH5 6NT, United Kingdom \\
  ${}^3$INAF -- Osservatorio Astronomico di Capodimonte, Salita Moiariello 16, 80131 Napoli, Italy \\
  ${}^4$Instituto de Astrofísica de Canarias, C/ Vía Láctea, s/n, E38205 - La Laguna (Tenerife). Spain}

\date{Accepted 2013 May 23. Received 2013 May 22; in original form 2013 April 26}

\pagerange{\pageref{firstpage}--\pageref{lastpage}} \pubyear{2013}

\begin{document}

\label{firstpage}

\maketitle

\begin{abstract}

We present orbit-based dynamical models and stellar population analysis of galaxy SDSS J151741.75-004217.6, a low-redshift ($z=0.116$) early-type galaxy (ETG) which, for its moderate luminosity, has an exceptionally high velocity dispersion. We aim to determine the central black hole mass ($\mbh$), the $i$-band stellar mass-to-light ratio ($\Upssti$), and the low-mass slope of the initial mass function (IMF). Combining constraints from HST imaging and long-slit kinematic data with those from fitting the SDSS spectrum with stellar populations  models of varying IMF, we show that this galaxy has a large fraction of low-mass stars, significantly higher than implied even by a Salpeter IMF. We exclude a Chabrier/Kroupa as well as a unimodal (i.e. single-segment) IMF, while a bimodal (low-mass tapered) shape is consistent with the dynamical constraints. Thereby, our study demonstrates that a very bottom-heavy IMF can exist even in an $L^\star$ ETG. We place an upper limit of $10^{10.5}\Msun$ on $\mbh$, which still leaves open the possibility of an extremely massive BH. 

\end{abstract}


%============================= section 1 =============================
\section{Introduction}
\label{sec:intro}

Recently, \cite{vdBosch+12} discovered a compact galaxy with very high central velocity dispersion ($\sigst$) that is caused by a high central mass density. Similar global properties are found in a number of objects presented in \cite{Bernardi+08} [B08 hereafter]. From their sample, SDSS J151741.75-004217.6 is an $L^\star$- galaxy ($4.7\times10^{10}\Lsun$) with effective radius of $R_e=1.5\kpc$ and $\sigst=360\kms$ \citep{Oh+11}. The combination of these properties is rare in the local universe \citep{Taylor+10}, and is a signature of either an over-massive central black hole (BH), or a high stellar mass-to-light ratio ($\Upsst$).

Based on analysis of high-definition spectroscopic data and up-to-date stellar spectral libraries, several recent publications \citep[e.g.][]{CvD12,Spinello+12} reported the existence of early-type galaxies (ETGs) with a large fraction of low-mass stars ($M\lesssim0.75\,\Msun$). Instead of being well represented by a \cite{Kroupa01} or \cite{Chabrier03} form, the IMF in these stellar systems appears to be bottom-heavy, as characterized by the IMF slope ($\Gamma$), which often exceeds even the slope of a \cite{Salpeter55} IMF ($\Gamma=1.35$). As a consequence, $\Upsst$ in such systems is higher than implied by the IMF found in the Local Group \citep[see e.g.][]{Bastian+10}, where it was observed directly based on number counts. Similar results for $\Upsst$ were also obtained in dynamical and lensing studies \citep{Thomas+11, Dutton+12}. In addition, $\Upsst$, $\Gamma$, and IMF-sensitive line strengths were found to correlate with several galaxy properties \citep[e.g.][]{CvD12}, most notably metallicity ($Z$) and $\sigst$ \citep[e.g.][]{Cenarro+03b,Cappellari+12,Ferreras+13}. These findings impact widely-used estimates of stellar mass ($\Mst$) from the observable galaxy magnitudes and colors (``SED fitting``). At the same time, the physical mechanisms determining the IMF shape, and its correlation with global galaxy properties, are currently unexplained. Probing the IMF jointly with galactic properties, especially in putative (local) outliers to these correlations, may offer valuable insights here.

To improve precision and systematic uncertainties when estimating $\Upsst(\Gamma)$, it is useful to supplement stellar population analysis with a dynamical analysis. Realistic dynamical models must include two non-luminous components: a dark matter (DM) halo and a central BH. The BH impacts the kinematics at small spatial scales (the BH gravitational \textit{sphere of influence}). Thus, if massive enough, the BH may also cause high central $\sigst$, rather than the stellar component. Current scientific consensus holds that BHs with masses $\mbh\gtrsim10^6\Msun$ are present in the centres of most massive galaxies, and the correlations of $\mbh$ with BH host galaxy properties have been well documented \citep[e.g.][]{FM00,Sani+11}. These \textit{BH scaling relations} are thought to be a product of the interplay between galaxy formation and BH growth. While several mechanisms have been suggested to explain their existence \citep[e.g.][]{SilkRees98, Croton+06, JahnkeMaccio11}, their empirical characterization continues to evolve and remains uncertain (e.g. \citealt{NFD06}; L\"asker et al. 2013, submitted). In particular, recent publications \citep{Rusli+11,vdBosch+12} reported BHs with $\mbh$ significantly exceeding the prediction of (some) BH scaling relations. If confirmed, these challenge the universality of the scaling relations and the associated BH-galaxy co-evolution models. It would therefore be desirable to ascertain if such \textit{\"{u}bermassive} BHs are rare outliers or instead represent a distinct BH population.

In order to estimate the stellar population properties, mass-to-light ratio, and BH mass of SDSS J151741.75-004217.6, we first construct fully general orbit-based dynamical models, based on HST imaging and ground-based long-slit kinematic data. Second, we use SDSS spectroscopic data and state-of-the-art stellar spectral libraries to analyse the stellar population and obtain an independent estimate of $\Upsst$. Finally, we compare and combine the results of both approaches. Throughout this letter, we adopt a $\Lambda$CDM cosmology with $(H_0,\Omega_M,\Omega_\Lambda)=(70\kms\Mpc^{-1},0.28,0.72)$.

%============================= section 2 =============================
\section{Data}
\label{sec:data}
%=====================================================================

SDSS J151741.75-004217.6 (2MASX J15174176-0042175), which we term ``b19'' for brevity, was listed in B08 as one of several galaxies with genuinely high velocity dispersion ($\sigst\gtrsim350\kms$), in which superposition effects could be excluded by means of HST ACS/HRC imaging. We chose b19 for further investigation because it is the intrinsically faintest and smallest object in B08's sample (object \#19 in their Table 2). The basic properties of b19 are listed in Table \ref{tab:b19_basic}. Given its small apparent size ($R_e=0\farcs9$), availability of a high-resolution image is crucial for modeling of the mass distribution. It is used in our modeling code in form of a multi-gaussian expansion (MGE, see Section \ref{sec:dynmod}).

\begin{figure}
  \begin{center}
    \includegraphics[width=\linewidth]{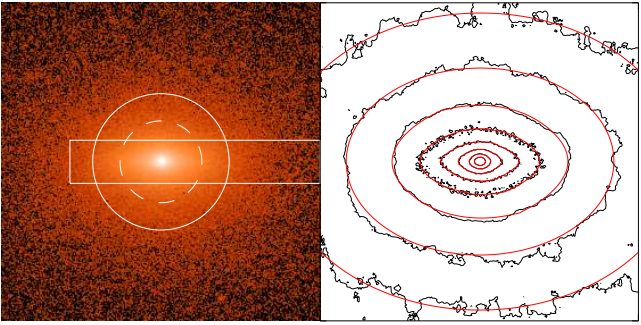}
  \end{center}
  \caption{Left panel: HST ACS/HRC-F775W ($i$-band) image of our target. Overlaid in solid lines are the kinmatic apertures (HET long-slit and SDSS fibre; see also Figure \ref{fig:b19_kinematics}), as well as the effective radius (dashed; $0\farcs9=1.9\kpc$). Right panel: surface brighntess contours of the image (black) in the range $[16,23]\magarcsec$, and those of the MGE model (Table \ref{tab:MGE_F775W}) overlaid in red. The displayed image area is $7\arcsec$ ($15\kpc$) across.}
  \label{fig:b19_image}
\end{figure}

We retrieved b19's $i$-band image (filter: F775W, PI: Bernardi, GO:10199) from the Hubble Legacy Archive. Its isophotes are regular and show no signs of dust lanes or spiral structure. We use \gfthree\ \citep{GF3} and the PSF obtained with TinyTim \citep{TinyTim11} to analyse the morphology and composition, and find that b19's radial surface brightness broadly follows a \sersic\ form with index $n=7$. However, the \sersic\ residual image shows substructure indicative of an embedded disk and a central cusp. A two-component model results in a highly flattened ($q=0.15$), faint (flux fraction $f_\mathrm{disk}\approx5\%$) and non-exponential ($n_\mathrm{disk}=1.6$) ``disk`` component. As for the single-component model, the \sersic\ index of the ''bulge`` is unusually high for its luminosity ($n_\mathrm{bul}=7.5$). Only by including a third \sersic\ component can we obtain a featureless residual image, and an exponential profile for the flattest component ($q=0.3$, 24\% flux fraction). Unfortunately, this model lacks a clear morphological interpretation for the two roundest components, which have $(R_e,n,q)=(0.26\kpc,2.0,0.61)$ and $(5.2\kpc,1.2,0.73)$ respectively. We therefore do not present a unique morphological classification, but surmise that b19 is a compact early-type galaxy that is dominated by a flattened spheroid and probably harbours an embedded, possibly non-exponential, disk. % \looseness=-1

The same image has been analyzed previously by \cite{Hyde+08}, whose results for magnitude and index of the single-\sersic\ profile agree well with ours. However, they derive a larger effective radius ($R_e=4.1\kpc$) than we do ($3.0\kpc$). Both are larger than the $R_e$ of a deVaucouleurs profile \citep[$1.8\kpc$,][]{Hyde+08}. Due to the apparent uncertainty in $R_e$, and the ambiguity in the decomposition, in the remainder of this letter we adopt $R_e$ and $i$-band luminosity ($L_i$) from circular aperture photometry (see Table \ref{tab:b19_basic}).

\begin{table}
\begin{tabular}{ccccc}
 \toprule
 quantity & value & method & ref. \\
 \midrule
 $z$ & 0.116 & SDSS & NED \\
 $d_A$ & $436\Mpc$ & $\Lambda$CDM & NED \\
 $d_L$ & $544\Mpc$ & $\Lambda$CDM & NED \\
 $(m-M)_L$ & 38.68 & -- & NED \\
 $a_i$ & 0.117 & SF11 & NED \\
 $L_i$  & $4.7\times10^{10}\Lsun$ & aper & -- \\
 $R_e$ & $0\farcs9$ / $1.9\kpc$ & aper & -- \\
 $\langle\mu_i\rangle_e-a_i$ & $18.24\magarcsec$ & aper & -- \\
 $n_\mathrm{ser}$ & 6.9 & Ser & -- \\
 $\sigst$ & $(360.3\pm9.4)\kms$ & SDSS & \cite{Oh+11} \\
 \bottomrule
\end{tabular}
\caption{Basic data for b19 (2MASX J15174176-0042175). Listed are (from top to bottom): redhsift, angular and luminosity distance, distance modulus, extinction, luminosity, effective radius, effective surface brightness, \sersic\ index, and central velocity dispersion. Photometric quantities are derived from the HST ACS/HRC-F775W image ($i$-band). Methods are ''SDSS``: based on the SDSS DR7 spectrum; ''SF11``: \protect\cite{SF11}, ''aper``: circular aperture photometry; and ''Ser``: 2D-\sersic\ profile.}
\label{tab:b19_basic}
\end{table}

\begin{figure}
  \begin{center}
    \includegraphics[width=\linewidth]{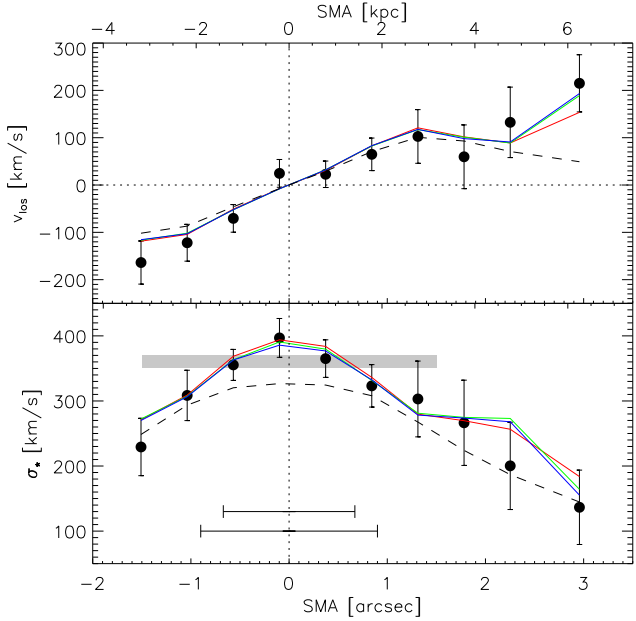}
  \end{center}
  \caption{Kinematic data of our target and selected orbit-based models. HET-based line-of-sight velocity ($\vlos$) and velocity dispersion ($\sigst$) are plotted as a function of position along the slit (semimajor axis), with the galaxy centre at the origin. The abscissa values represent the flux-weighted average position of each spectral bin, and error bars the $1\sigma$ measurement uncertainties. The SDSS aperture, and $\sigst$ ($1\sigma$-confidence) is indicated by the gray rectangle. Solid (red, green, blue) curves and the dashed curve are the best-fit models with $(\Upssti,\log\mbh)=(2.6,10.3)$, $(6.0,10.3)$, $(6.0,8.6)$, and $(2.6,8.3)$. The FWHM of the HET (SDSS) PSF is displayed by the upper (lower) horizontal bar.}
  \label{fig:b19_kinematics}
\end{figure}

Kinematic data for our target have been obtained on the LRS \citep{HET-LRS98} at the Hobby-Eberly Telescope (HET); its reduction is described in \cite{vdBosch+12} and van den Bosch et al. (2013, in prep.). The $0\farcs95$-wide long-slit has a pixel scale of $0\farcs475$, and the spectral resolution is $4.5\text{\AA}$. We reconstruct its spatial PSF ($\text{FWHM}=1\farcs34$) as a sum of two Gaussian components with (weight,dispersion) of $(0.2,1\farcs11)$ and $(0.8,0\farcs54)$, used to convolve the projected kinematics of the dynamical models before comparing with the data (Section \ref{sec:dynmod}). We use pPXF \citep{pPXF} and MILES template spectra \citep{MILES11} to measure the line-of-sight velocity and velocity dispersion ($\vlos$ and $\sigst$) out to a central distance of $\sim3\arcsec$ ($\sim1.6\,R_e$) along the major axis (see Figure \ref{fig:b19_kinematics}). In addition to a high central velocity dispersion peak of $\approx400\kms$, we detect significant rotation around the photometric minor axis, which strongly implies an intrinsic oblate geometry, as opposed to triaxial or prolate. Our long-slit data are supplemented by, and consistent with, $\sigst$ as measured in the SDSS fibre (diameter $3\arcsec$, PSF $\text{FWHM}=1\farcs8$) by \cite{Oh+11}.
 
%=====================================================================
\section{Dynamical modeling}
 \label{sec:dynmod}

We use the HST imaging and SDSS+HET kinematic data to constrain dynamical models that have been constructed via the method of \cite{Schwarzschild79}. Based on a comprehensive library of representative stellar orbits, this method allows for maximal freedom in orbital structure that may be especially important here considering the peculiar properties of b19. It thereby has an advantage over other methods, such as fitting solutions of the (axisymmetric) Jeans equations or analytic distribution functions. We explore only models with oblate intrinsic shape as implied by the observed rotational pattern.

Our implementation of the Schwarzschild method is described in \cite{vdBosch+08_triaxSSM}. It uses an MGE parametrization of the surface brightness profile \citep{Emsellem+94,Cappellari_MGE}, which allows for an analytic deprojection. We use \gfthree\ to derive the MGE parameters (see Table \ref{tab:MGE_F775W} and right panel of Figure \ref{fig:b19_image}). It was necessary to include a quite flattened ($q=0.26$) component for an acceptable fit, which probably represents an embedded disk and effectively limits the range of inclination ($i$) to $77^\circ\lesssim i \lesssim 90^\circ$.

\begin{table}
 \centering
  \begin{tabular}{*{3}{c}}
   \toprule
   $m_j$ & $\sigma_j\,['']$ & $q_j$ \\
   \midrule
    % m_i & sig_i & q_i
19.81 & 0.0329 & 0.491 \\
19.11 & 0.0957 & 0.647 \\
19.68 &  0.378 & 0.256 \\
19.26 &  0.220 & 0.687 \\
18.49 &  0.660 & 0.439 \\
18.39 &   1.28 & 0.673 \\
17.79 &   2.95 & 0.748 \\

   \bottomrule
  \end{tabular}
  \caption{MGE parametrization of our target in the $i$-band (HST/ACS F775W), used as input to the dynamical models. $m_j$ is the apparent magnitude of the $j$-th Gaussian component (before extinction correction), $\sigma_j$ its dispersion, and $q_j$ its axis ratio.}
  \label{tab:MGE_F775W}
\end{table} 

We run models assuming minimal and  maximal allowed inclination to deproject the surface brightness. The resulting luminosity density is converted to the stellar mass density via the global stellar mass-to-light ratio in the $i$-band, ${\Upsst}_{,i}\equiv\Mst/L_i$. Given the spatial resolution of our kinematic data, we therefore ignore possible gradients in $\Upsst$. Further gravitational sources in our models are a central BH, characterized by $\mbh$, and a dark matter (DM) halo of virial mass $\Mdm \equiv M_{200}$. We define $f_\mathrm{DM}=\Mdm/\Mst$,  the ratio between $\Mdm$ and stellar mass $\Mst$, and vary it in the range $[1,100]$. The model DM halo is spherically symmetric with an NFW profile \citep[][]{NFW97}. Given the previously estimated $\Mst$ of $1.15\times10^{11}\Msun$ (MPA-JHU DR7\footnote{http://www.mpa-garching.mpg.de/SDSS/DR7/Data/stellarmass.html}), we expect $\Mdm\approx2\times10^{12}\Msun$ \citep{Moster+10}, and thus a halo concentration index of $c_{200}=8$ \citep{MaccioDuttonBosch08}, which we adopt throughout. Theoretically, $c$ is anticorrelated with $\Mdm$, varying between $[5,12]$ over range $\Mdm\in[0.5,500]\times10^{11}\Msun$ of the model grid. Yet, its exact value should be insignificant, as is $\Mdm$, because the halo scale radius is $\sim10-100$ times larger than $R_e$, and the DM density is very low compared to the baryonic density ($\approx1\%$ inside $1\,R_e$ even for the most massive haloes). For the same reason, our results should be insensitive to the central DM density log-slope, i.e. possible departures from an NFW form.

\begin{figure}
  \begin{center}
    \includegraphics[width=\linewidth]{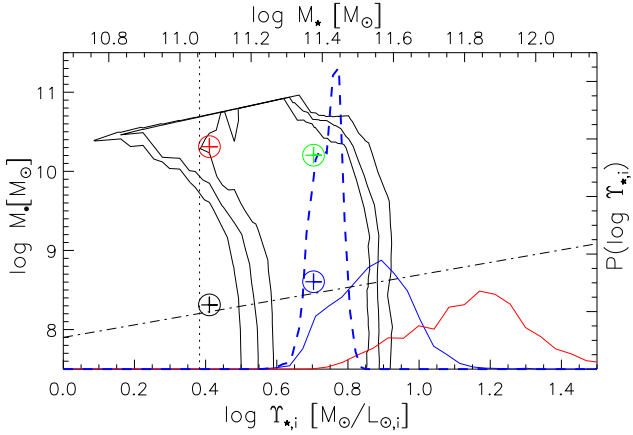}
  \end{center}
  \caption{Black contours: $\chi^2\,(\log\Upssti,\log\mbh)$ of dynamic models fitted to b19's imaging and kinematic data, with galaxy mass indicated by the upper x-axis. DM fraction and inclination have been marginalized over. Shown levels are the $\{68.3\%,95.5\%,99.7\%\}$-quantiles of a $\chi^2$-distribution with two degrees of freedom. Kinematics of select models (symbols) are plotted in Figure \ref{fig:b19_kinematics}. Blue/red solid curves: probability density of $\log\Upssti$ from our stellar population analysis, where the slope of the IMF with bimodal/unimodal shape was a free parameter. The blue dashed curve is the joint probability from both methods when adopting a bimodal IMF. For comparison, the dotted and dot-dashed lines respectively represent $\Mst$ implied by a Chabrier IMF (MPA-JHU DR7 data release$^1$), and the $\mbh-\Mst$ relation \protect\citep{Sani+11}. For details, see Sections \ref{sec:dynmod} and \ref{sec:popana}.}
  \label{fig:mainfig}
\end{figure}

In total, we compute dynamical models on a $37\times25\times11\times2$-grid evenly spaced in parameters $\Upssti\in[1,10]$, $\log(\mbh/\Msun)\in[7.5,11.5]$, $\log(f_\mathrm{DM})\in[0,2]$ and $i\in\{77^\circ,90^\circ\}$. Using the $\chi^2$-statistic, the orbital weights of each model are optimized to fit the photometric and kinematic data. In order to evaluate the relative likelihoods of the parameters of interest, we marginalize the models over $f_\mathrm{DM}$ and $i$, and plot the minimum-$\chi^2$ as a function of $(\log\Upssti,\log\mbh)$. Figure \ref{fig:mainfig} presents contours (black solid) of $\Delta\chi^2=\{2.3,\,6.2,\,11.8\}$, corresponding to the $\{68.3\%,95.5\%,99.7\%\}$-quantiles of the $\chi^2$-distribution with two degrees of freedom. Models with $\log\mbh\approx 10.3$ fit the data well across a wide range of assumed mass-to-light ratios, as do models with $0.6\lesssim{\log\Upsst}_{,i}\lesssim 0.85$ when $\mbh\lesssim10^{10}\Msun$. Our data are not able to further distinguish between models in this parameter range due to limited depth and resolution, which prevents spatially resolved measurement of higher-order velocity moments and leaves the orbital anisotropy in the central parts (the BH sphere-of-influence) relatively unconstrained. We verified that all our models are insensitive to the presence of dark matter, a direct consequence of b19's small size and high stellar mass density ($\approx5\,\Msun/\mathrm{pc}^3$ inside $R_e$ if $\log\Upsst\approx0.8$). The quality of the fit is illustrated in Figure \ref{fig:b19_kinematics}, where selected models are compared to the kinematic data. Models  with $\log(\mbh/\Msun)\approx10.3$ or $\log\Upssti=0.7$ (solid coloured lines) fit the data very well and differ only insignificantly in their observables, while models with $\log\mbh\lesssim9.5$ and $\log\Upssti\lesssim0.55$ (dashed line, corresponding to the black cross in the top panel of Figure \ref{fig:mainfig}) are strongly disfavoured.

This implies that, if $\log\Upssti=0.4$ as estimated for b19 by stellar population synthesis and assuming a Chabrier IMF (MPA-JHU DR7 data release$^1$, dotted line in Figure \ref{fig:mainfig}), $\log(\mbh/\Msun)\in[10.2,10.6]$ at the $68\%$-confidence level. This is 2 orders of magnitude, or 5 standard deviations, above the $\mbh-\Mst$-relation between BH and host galaxy bulge mass \citep[][dash-dotted line]{Sani+11}. Conversely, if b19 is assumed to follow $\mbh-\Mst$, the dynamical models require $\log\Upssti\in[0.60,0.84]$, i.e. about twice the stellar mass expected from a Chabrier/Kroupa IMF.

\section{Stellar Population and IMF}
\label{sec:popana}

To narrow the parameter space (\"{u}bermassive BH or bottom-heavy IMF) allowed by the photometric and kinematic data, we obtain an independent $\Upssti$ estimate from the SDSS spectrum of b19, by fitting the line strengths of gravity-sensitive features with state-of-the-art extended-MILES (MIUSCAT, \citealt{MIUSCAT}) stellar population models. Our applied method is identical to that of \cite{Ferreras+13}. In this ''hybrid approach``, we consider SSPs as well as models with an exponentially declining SFR. This is an advantage with respect to the standalone line strength fits where, so far, we have considered only single SSPs. In addition to fits to the full spectrum in the range $3400\text{\AA}<\lambda<5400\text{\AA}$, we fit equivalent widths (EWs) of spectral absorption lines that are sensitive to age (H$\beta_0$, H$\gamma_F$), metallicity $[Z/H]$ ($[\text{MgFe}]$) and IMF slope $\Gamma$ (TiO$_1$, TiO$_2$, NaD and Na8190). Using these indices, estimates of age and $\Gamma$ are robust against non-solar abundance ratios \citep[see][]{LaBarbera+13}. We consider two different IMF shapes: ''unimodal'' (single powerlaw) and ''bimodal'' (Kroupa-like, i.e. tapered at low mass) as defined in \cite{Vazdekis+96}, with $\Gamma$ being a free parameter. Additionally, we take intragalactic dust reddening, $E(B-V)$, into account. Our models encompass parameter ranges $[Z/H]\in[-1.0,0.22]$, $\Gamma\in[0.3,3.3]$, $E(B-V)\in[0,1]$, and ages between 1-13 Gyr. We compute the joint likelihood from the fit to the spectrum and EWs, and marginalize to obtain the probability distribution (PDF) of each parameter separately. \looseness=-1

This analysis shows that b19 is composed of an old stellar population, and that internal dust extinction is very low (mass-weighted age $>8.5\Gyr$ and $E(B-V)<0.1$ at 90\% confidence). From the measured EWs, we also find strong indications for a high (super-solar) metallicity and $\alpha$-abundance ($[\alpha/\text{Fe}]$), suggesting an intense, short-lived star formation process, as expected from an object with such a high velocity dispersion \citep{deLaRosa+11}. All IMF-sensitive line strengths indicate a large fraction of low-mass stars, with $\Gamma=3.2\pm0.1\,(2.4\pm0.3)$ for a bimodal (unimodal) IMF shape. For comparison, a Kroupa/Chabrier (Salpeter) IMF has $\Gamma=1.3\,(1.35)$. The resulting PDFs of $\Upssti$ are plotted Figure \ref{fig:mainfig} by the blue (red) curves. The results for the bimodal IMF, $\log{\Upssti}=0.85\pm0.11$ ($1\sigma$-uncertainties), is consistent with the dynamical models, while the unimodal IMF implies significantly larger values ($1.12\pm0.16$). We note that a log-normal IMF would probably also be consistent with the data when using a lower turnover mass and a steeper high-mass slope than the Chabrier IMF (e.g. $0.2\,\Msun$ and $\Gamma=2.1$ for the same $<0.5\Msun$ fraction as our best-fitting bimodal IMF). Adopting the bimodal shape, and combining the resulting PDF in $\log\Upssti$ with the constraints from dynamical modeling after marginalizing over $\mbh$, we determine that $\log\Upssti\in[0.60,0.82]$ at the $3\sigma$-confidence level, up to $\approx160\,(60)\%$ larger than for a Chabrier/Kroupa (Salpeter) IMF. \looseness=-1

\section{Discussion}

Dynamical models of low-resolution long-slit and single-fibre kinematic data, in combination with high-resolution HST imaging, affirmed that b19 either hosts an extremely massive BH ($\mbh\approx2\times10^{10}$ for the given luminosity of $L_i=4.7\times10^{10}\Lsun$), or has an unexpectedly high stellar mass-to-light ratio. Resolution and depth of the kinematic data was not yet sufficient to assign preference to either scenario. In order to add an independent constraint on $\Upsst$, we modeled the spectrum with state-of-the-art \miuscat\ SSP stellar population models with a variety of SFHs and IMFs, accounting for spectral features sensitive to the fraction of dwarf stars that was previously unconstrained in distant (unresolved) stellar systems. In the present case, the spectra unambiguously indicate a very ''bottom-heavy`` IMF, with a slope well in excess of a Kroupa- and Salpeter-form. At the implied high stellar mass-to-light ratio, $\mbh$ is dynamically not constrained to low values, but has an upper limit of $\approx10^{10.5}\Msun$. If instead b19 had a ''normal'' IMF, the dynamics would dictate an \"ubermassive BH ($\approx10^{10}\Msun$).

Conversely, since our dynamical modeling constrains $\Upssti$, it indicates that the IMF shape is most likely not unimodal, but rather bimodal. While the spectral lines of low-mass stars can be satisfactorily fit using either of both functional forms (albeit with a different slope, respectively), the resulting $\Upssti$ differs by a factor of $\approx 2$. Therefore, our analysis demonstrates that stellar kinematics can serve as a valuable tool to constrain the shape of the IMF at the low-mass end.

Galaxy b19 is kinematically and morphologically quite distinct from the vast majority of present-day ETGs. It is substantially flattened and, considering $v/\sigst\sim0.5$, a ''fast rotator'' \citep{SAURON-IX}. Yet, in constrast to these, b19 exhibits an extremely high $\sigst$ and is a clear outlier in the fundamental plane of ellipticals: following \cite{Bernardi+03}, $R_e(\sigst,\langle\mu_i\rangle_e)=(6.9\pm1.6)\kpc$ instead of the observed $1.9\kpc$. Its compactness, morphology and kinematics make b19 appear similar to the compact ETGs abundant at redshift $1 \lesssim z \lesssim 3$ \citep[e.g.,][]{vDokkum+08b,vdWel+08b,vdWel+11}, which have grown significantly in size since \citep[e.g.][]{Trujillo+11}. A prominent mechanism proposed to drive this evolution is repeated dry merging, especially minor merging, in the hierarchy of concordance cosmology \citep[e.g.][]{HilzNaabOstriker13}. In this picture, b19 would be a ''survivor`` of the dominant population at high redshift, having somehow avoided the typical evolutionary path of relatively quiescent size increase, and thus represents an earlier stage of galaxy formation. This interpretation is also supported by b19's stellar age, $\alpha$-abundance, and the absence of dust. \looseness=-1

At the same time, b19 appears to follow the \textit{local} correlations between $\sigst$ and stellar population properties, ($Z$, $[\alpha/\text{Fe}]$, and $\Upsst(\Gamma)$ in particular). Supposing they were already in place before the onset of significant minor and dry merging, these correlations should have become shallower and weaker since. However, if indeed ETGs are assembled inside-out, the present-day correlation of $\Upsst$ with $\sigst$ may be partially preserved in the galaxy centres and still reflect the connection between SF physics and global conditions (halo concentration, gas density) that early-on led to a bottom-heavy IMF as found in b19. 

It is therefore worthwhile to study the IMF in more objects with properties similar to b19. Together with measuring potential IMF gradients, this investigation could help illuminating the physics of SF, as well as galaxy evolution at large. To better estimate $\Upsst\,(\Gamma)$ and $\mbh$, dynamical or lensing models should be included in the analysis and constrained by high-resolution IFU spectroscopy (e.g. VLT SINFONI): for example in NGC1277, ${\Upsst}_{,V}=(6\pm4)\Mst/{\Lsun}_{,V}$ ($3\sigma$-level) is relatively uncertain (and a bottom-heavy IMF cannot be ruled out), while $\mbh$ is dynamically well constrained at $\sim10^{9.5\dots10.5}\Msun$ and only weakly correlated with $\Upsst$ (see figure S1 in \citealt{vdBosch+12}; \citealt{Emsellem13}). In case of b19, improved data could determine if a $\sim10^{10}\Msun$ BH is present, which is an important advantage considering that central BHs are also thought to play a crucial role in galaxy formation. % \looseness=-1
%=====================================================================

\section*{Acknowledgments}
\label{sec:acknowledgments}

We thank Jarle Brinchmann, Stephen Bailey and David Schlegel for helpful information regarding MPA-JHU derivation of stellar masses and the seeing characterization of the SDSS spectra, as well as Arjen van der Wel for helpful discussions and Laura Watkins for a review of the manuscript. RL acknowledges support by the Heidelberg Graduate School of Fundamental Physics. The Hobby-Eberly Telescope is a joint project of the University of Texas at Austin, the Pennsylvania State University, Ludwig-Maximilians-Universit\"{a}t M\"{u}nchen, and Georg-August-Universit\"{a}t G\"{o}ttingen. It is named in honor of its principal benefactors, William P. Hobby and Robert E. Eberly.

%=====================================================================
% REFERENCES
%=====================================================================

\bibliographystyle{mn2e}
\bibliography{B19}

\begin{thebibliography}{48}
\expandafter\ifx\csname natexlab\endcsname\relax\def\natexlab#1{#1}\fi

\bibitem[{{Bastian}, {Covey} \& {Meyer}(2010){Bastian}, {Covey}, \&
  {Meyer}}]{Bastian+10}
{Bastian} N., {Covey} K.~R., {Meyer} M.~R., 2010, \araa, 48, 339

\bibitem[{{Bernardi} {et~al}\mbox{.}(2008){Bernardi}, {Hyde}, {Fritz}, {Sheth},
  {Gebhardt}, \& {Nichol}}]{Bernardi+08}
{Bernardi} M., {Hyde} J.~B., {Fritz} A., {Sheth} R.~K., {Gebhardt} K., {Nichol}
  R.~C., 2008, \mnras, 391, 1191

\bibitem[{{Bernardi} {et~al}\mbox{.}(2003){Bernardi}, {Sheth}, {Annis},
  {Burles}, {Eisenstein}, {Finkbeiner}, {Hogg}, {Lupton}, {Schlegel},
  {SubbaRao}, {Bahcall}, {Blakeslee}, {Brinkmann}, {Castander}, {Connolly},
  {Csabai}, {Doi}, {Fukugita}, {Frieman}, {Heckman}, {Hennessy}, {Ivezi{\'c}},
  {Knapp}, {Lamb}, {McKay}, {Munn}, {Nichol}, {Okamura}, {Schneider}, {Thakar},
  \& {York}}]{Bernardi+03}
{Bernardi} M. {et~al.}, 2003, \aj, 125, 1866

\bibitem[{{Cappellari}(2002)}]{Cappellari_MGE}
{Cappellari} M., 2002, \mnras, 333, 400

\bibitem[{{Cappellari} \& {Emsellem}(2004)}]{pPXF}
{Cappellari} M., {Emsellem} E., 2004, \pasp, 116, 138

\bibitem[{{Cappellari} {et~al}\mbox{.}(2012){Cappellari}, {McDermid},
  {Alatalo}, {Blitz}, {Bois}, {Bournaud}, {Bureau}, {Crocker}, {Davies},
  {Davis}, {de Zeeuw}, {Duc}, {Emsellem}, {Khochfar}, {Krajnovi{\'c}},
  {Kuntschner}, {Lablanche}, {Morganti}, {Naab}, {Oosterloo}, {Sarzi}, {Scott},
  {Serra}, {Weijmans}, \& {Young}}]{Cappellari+12}
{Cappellari} M. {et~al.}, 2012, \nat, 484, 485

\bibitem[{{Cenarro} {et~al}\mbox{.}(2003){Cenarro}, {Gorgas}, {Vazdekis},
  {Cardiel}, \& {Peletier}}]{Cenarro+03b}
{Cenarro} A.~J., {Gorgas} J., {Vazdekis} A., {Cardiel} N., {Peletier} R.~F.,
  2003, \mnras, 339, L12

\bibitem[{{Chabrier}(2003)}]{Chabrier03}
{Chabrier} G., 2003, \apjl, 586, L133

\bibitem[{{Conroy} \& {van Dokkum}(2012)}]{CvD12}
{Conroy} C., {van Dokkum} P.~G., 2012, \apj, 760, 71

\bibitem[{{Croton} {et~al}\mbox{.}(2006){Croton}, {Springel}, {White}, {De
  Lucia}, {Frenk}, {Gao}, {Jenkins}, {Kauffmann}, {Navarro}, \&
  {Yoshida}}]{Croton+06}
{Croton} D.~J. {et~al.}, 2006, \mnras, 365, 11

\bibitem[{{de La Rosa} {et~al}\mbox{.}(2011){de La Rosa}, {La Barbera},
  {Ferreras}, \& {de Carvalho}}]{deLaRosa+11}
{de La Rosa} I.~G., {La Barbera} F., {Ferreras} I., {de Carvalho} R.~R., 2011,
  \mnras, 418, L74

\bibitem[{{Dutton}, {Mendel} \& {Simard}(2012){Dutton}, {Mendel}, \&
  {Simard}}]{Dutton+12}
{Dutton} A.~A., {Mendel} J.~T., {Simard} L., 2012, \mnras, 422, L33

\bibitem[{{Emsellem}(2013)}]{Emsellem13}
{Emsellem} E., 2013, accepted to \mnras (arXiv:1305.3630)

\bibitem[{{Emsellem} {et~al}\mbox{.}(2007){Emsellem}, {Cappellari},
  {Krajnovi{\'c}}, {van de Ven}, {Bacon}, {Bureau}, {Davies}, {de Zeeuw},
  {Falc{\'o}n-Barroso}, {Kuntschner}, {McDermid}, {Peletier}, \&
  {Sarzi}}]{SAURON-IX}
{Emsellem} E. {et~al.}, 2007, \mnras, 379, 401

\bibitem[{{Emsellem}, {Monnet} \& {Bacon}(1994){Emsellem}, {Monnet}, \&
  {Bacon}}]{Emsellem+94}
{Emsellem} E., {Monnet} G., {Bacon} R., 1994, \aap, 285, 723

\bibitem[{{Falc{\'o}n-Barroso}(2011)}]{MILES11}
{Falc{\'o}n-Barroso} J. e.~a., 2011, \aap, 532, A95

\bibitem[{{Ferrarese} \& {Merritt}(2000)}]{FM00}
{Ferrarese} L., {Merritt} D., 2000, \apjl, 539, L9

\bibitem[{{Ferreras} {et~al}\mbox{.}(2013){Ferreras}, {La Barbera}, {de la
  Rosa}, {Vazdekis}, {de Carvalho}, {Falc{\'o}n-Barroso}, \&
  {Ricciardelli}}]{Ferreras+13}
{Ferreras} I., {La Barbera} F., {de la Rosa} I.~G., {Vazdekis} A., {de
  Carvalho} R.~R., {Falc{\'o}n-Barroso} J., {Ricciardelli} E., 2013, \mnras,
  429, L15

\bibitem[{{Hill}(1998)}]{HET-LRS98}
{Hill} G.~J. e.~a., 1998, in SPIE, Vol. 3355, pp. 433--443

\bibitem[{{Hilz}, {Naab} \& {Ostriker}(2013){Hilz}, {Naab}, \&
  {Ostriker}}]{HilzNaabOstriker13}
{Hilz} M., {Naab} T., {Ostriker} J.~P., 2013, \mnras, 429, 2924

\bibitem[{{Hyde} {et~al}\mbox{.}(2008){Hyde}, {Bernardi}, {Sheth}, \&
  {Nichol}}]{Hyde+08}
{Hyde} J.~B., {Bernardi} M., {Sheth} R.~K., {Nichol} R.~C., 2008, \mnras, 391,
  1559

\bibitem[{{Jahnke} \& {Macci{\`o}}(2011)}]{JahnkeMaccio11}
{Jahnke} K., {Macci{\`o}} A.~V., 2011, \apj, 734, 92

\bibitem[{{Krist}, {Hook} \& {Stoehr}(2011){Krist}, {Hook}, \&
  {Stoehr}}]{TinyTim11}
{Krist} J.~E., {Hook} R.~N., {Stoehr} F., 2011, in SPIE, Vol. 8127

\bibitem[{{Kroupa}(2001)}]{Kroupa01}
{Kroupa} P., 2001, \mnras, 322, 231

\bibitem[{{La Barbera}(2013)}]{LaBarbera+13}
{La Barbera}, F. e.~a., 2013, \mnras, submitted (arXiv:1305.2273)

\bibitem[{{Macci{\`o}}, {Dutton} \& {van den Bosch}(2008){Macci{\`o}},
  {Dutton}, \& {van den Bosch}}]{MaccioDuttonBosch08}
{Macci{\`o}} A.~V., {Dutton} A.~A., {van den Bosch} F.~C., 2008, \mnras, 391,
  1940

\bibitem[{{Moster} {et~al}\mbox{.}(2010){Moster}, {Somerville}, {Maulbetsch},
  {van den Bosch}, {Macci{\`o}}, {Naab}, \& {Oser}}]{Moster+10}
{Moster} B.~P., {Somerville} R.~S., {Maulbetsch} C., {van den Bosch} F.~C.,
  {Macci{\`o}} A.~V., {Naab} T., {Oser} L., 2010, \apj, 710, 903

\bibitem[{{Navarro}, {Frenk} \& {White}(1997){Navarro}, {Frenk}, \&
  {White}}]{NFW97}
{Navarro} J.~F., {Frenk} C.~S., {White} S.~D.~M., 1997, \apj, 490, 493

\bibitem[{{Novak}, {Faber} \& {Dekel}(2006){Novak}, {Faber}, \&
  {Dekel}}]{NFD06}
{Novak} G.~S., {Faber} S.~M., {Dekel} A., 2006, \apj, 637, 96

\bibitem[{{Oh} {et~al}\mbox{.}(2011){Oh}, {Sarzi}, {Schawinski}, \&
  {Yi}}]{Oh+11}
{Oh} K., {Sarzi} M., {Schawinski} K., {Yi} S.~K., 2011, \apjs, 195, 13

\bibitem[{{Peng} {et~al}\mbox{.}(2010){Peng}, {Ho}, {Impey}, \& {Rix}}]{GF3}
{Peng} C.~Y., {Ho} L.~C., {Impey} C.~D., {Rix} H.-W., 2010, \aj, 139, 2097

\bibitem[{{Rusli} {et~al}\mbox{.}(2011){Rusli}, {Thomas}, {Erwin}, {Saglia},
  {Nowak}, \& {Bender}}]{Rusli+11}
{Rusli} S.~P., {Thomas} J., {Erwin} P., {Saglia} R.~P., {Nowak} N., {Bender}
  R., 2011, \mnras, 410, 1223

\bibitem[{{Salpeter}(1955)}]{Salpeter55}
{Salpeter} E.~E., 1955, \apj, 121, 161

\bibitem[{{Sani} {et~al}\mbox{.}(2011){Sani}, {Marconi}, {Hunt}, \&
  {Risaliti}}]{Sani+11}
{Sani} E., {Marconi} A., {Hunt} L.~K., {Risaliti} G., 2011, \mnras, 413, 1479

\bibitem[{{Schlafly} \& {Finkbeiner}(2011)}]{SF11}
{Schlafly} E.~F., {Finkbeiner} D.~P., 2011, \apj, 737, 103

\bibitem[{{Schwarzschild}(1979)}]{Schwarzschild79}
{Schwarzschild} M., 1979, \apj, 232, 236

\bibitem[{{Silk} \& {Rees}(1998)}]{SilkRees98}
{Silk} J., {Rees} M.~J., 1998, \aap, 331, L1

\bibitem[{{Spiniello} {et~al}\mbox{.}(2012){Spiniello}, {Trager}, {Koopmans},
  \& {Chen}}]{Spinello+12}
{Spiniello} C., {Trager} S.~C., {Koopmans} L.~V.~E., {Chen} Y.~P., 2012, \apjl,
  753, L32

\bibitem[{{Taylor} {et~al}\mbox{.}(2010){Taylor}, {Franx}, {Glazebrook},
  {Brinchmann}, {van der Wel}, \& {van Dokkum}}]{Taylor+10}
{Taylor} E.~N., {Franx} M., {Glazebrook} K., {Brinchmann} J., {van der Wel} A.,
  {van Dokkum} P.~G., 2010, \apj, 720, 723

\bibitem[{{Thomas} {et~al}\mbox{.}(2011){Thomas}, {Saglia}, {Bender}, {Thomas},
  {Gebhardt}, {Magorrian}, {Corsini}, {Wegner}, \& {Seitz}}]{Thomas+11}
{Thomas} J. {et~al.}, 2011, \mnras, 415, 545

\bibitem[{{Trujillo}, {Ferreras} \& {de La Rosa}(2011){Trujillo}, {Ferreras},
  \& {de La Rosa}}]{Trujillo+11}
{Trujillo} I., {Ferreras} I., {de La Rosa} I.~G., 2011, \mnras, 415, 3903

\bibitem[{{van den Bosch} {et~al}\mbox{.}(2012){van den Bosch}, {Gebhardt},
  {G{\"u}ltekin}, {van de Ven}, {van der Wel}, \& {Walsh}}]{vdBosch+12}
{van den Bosch} R.~C.~E., {Gebhardt} K., {G{\"u}ltekin} K., {van de Ven} G.,
  {van der Wel} A., {Walsh} J.~L., 2012, \nat, 491, 729

\bibitem[{{van den Bosch} {et~al}\mbox{.}(2008){van den Bosch}, {van de Ven},
  {Verolme}, {Cappellari}, \& {de Zeeuw}}]{vdBosch+08_triaxSSM}
{van den Bosch} R.~C.~E., {van de Ven} G., {Verolme} E.~K., {Cappellari} M.,
  {de Zeeuw} P.~T., 2008, \mnras, 385, 647

\bibitem[{{van der Wel} {et~al}\mbox{.}(2011){van der Wel}, {Rix}, {Wuyts},
  {McGrath}, {Koekemoer}, {Bell}, {Holden}, {Robaina}, \&
  {McIntosh}}]{vdWel+11}
{van der Wel} A. {et~al.}, 2011, \apj, 730, 38

\bibitem[{{van der Wel} \& {van der Marel}(2008)}]{vdWel+08b}
{van der Wel} A., {van der Marel} R.~P., 2008, \apj, 684, 260

\bibitem[{{van Dokkum} {et~al}\mbox{.}(2008){van Dokkum}, {Franx}, {Kriek},
  {Holden}, {Illingworth}, {Magee}, {Bouwens}, {Marchesini}, {Quadri},
  {Rudnick}, {Taylor}, \& {Toft}}]{vDokkum+08b}
{van Dokkum} P.~G. {et~al.}, 2008, \apjl, 677, L5

\bibitem[{{Vazdekis} {et~al}\mbox{.}(1996){Vazdekis}, {Casuso}, {Peletier}, \&
  {Beckman}}]{Vazdekis+96}
{Vazdekis} A., {Casuso} E., {Peletier} R.~F., {Beckman} J.~E., 1996, \apjs,
  106, 307

\bibitem[{{Vazdekis} {et~al}\mbox{.}(2012){Vazdekis}, {Ricciardelli},
  {Cenarro}, {Rivero-Gonz{\'a}lez}, {D{\'{\i}}az-Garc{\'{\i}}a}, \&
  {Falc{\'o}n-Barroso}}]{MIUSCAT}
{Vazdekis} A., {Ricciardelli} E., {Cenarro} A.~J., {Rivero-Gonz{\'a}lez} J.~G.,
  {D{\'{\i}}az-Garc{\'{\i}}a} L.~A., {Falc{\'o}n-Barroso} J., 2012, \mnras,
  424, 157

\end{thebibliography}

%=====================================================================
% APPENDICES
%=====================================================================

\label{lastpage}

\end{document}